\documentclass[letterpaper, 10 pt, conference]{ieeeconf}  % Comment this line out if you need a4paper

\IEEEoverridecommandlockouts                              % This command is only needed if 
\usepackage{amssymb}
\usepackage{amsmath}
\usepackage{adjustbox}
\usepackage{subfigure}
\usepackage{algorithm}
\usepackage{enumerate}
\usepackage{booktabs}
\usepackage{makecell}
\usepackage{multirow}
\usepackage[capitalize]{cleveref}
\usepackage{svg}
\crefname{section}{Sec.}{Secs.}
\Crefname{table}{Table}{Tables}

\newtheorem{asm}{Assumption}

\graphicspath{{fig/}}
\title{\LARGE \bf Co-optimization of Vehicle Dynamics and Powertrain Management for Connected and Automated Electric Vehicles}
\author{Zongtan Li, Yunli Shao
\thanks{Zongtan Li, and Yunli Shao (corresponding author, Email:
yunli.shao@uga.edu) are with the School of Environmental, Civil, Agricultural and Mechanical Engineering, University of Georgia, Athens, 30602}%
}

\begin{document}

\maketitle
\thispagestyle{empty}
\pagestyle{empty}
% Zongtan Li Pin: 222901 117657
\begin{abstract}
Connected and automated vehicles (CAVs) represent the future of transportation, utilizing detailed traffic information to enhance control and decision-making. Eco-driving of CAVs has the potential to significantly improve energy efficiency, and the benefits are maximized when both vehicle speed and powertrain operation are optimized. In this paper, we studied the co-optimization of vehicle speed and powertrain management for energy savings in a dual-motor electric vehicle. Control-oriented vehicle dynamics and electric powertrain models were developed to transform the problem into an optimal control problem specifically designed to facilitate real-time computation. Simulation validation was conducted using real-world data calibrated traffic simulation scenarios in Chattanooga, TN. Evaluation results demonstrated a 12.80-24.52\% reduction in the vehicle's power consumption under ideal predicted traffic conditions, while maintaining benefits with various prediction uncertainties, such as Gaussian process uncertainties on acceleration and time-shift effects on predicted speed. The energy savings of the proposed eco-driving strategy are achieved through effective speed control and optimized torque allocation. The proposed model can be extended to various CAV and electric vehicle applications, with potential adaptability to diverse traffic scenarios.
\end{abstract}
    %  The proposed method is validated by simulation in VISSIM and hardware in the loop (HIL) test. The results show that the proposed method can save energy and improve the vehicle's performance.
% \keywords  Power-train modeling, Connected Autonomous Vehicle, eco-driving  \endkeywords
  
\section{Introduction}
Reducing carbon and greenhouse gas emissions plays a crucial role in addressing environmental changes. The U.S. has set a goal to cut greenhouse gas emissions by 50-52\% from 2005 levels by 2030. Additionally, according to data from the Department of Energy ~\cite{DOE2020}, eco-driving can significantly improve energy efficiency and holds significant commercialization potential. Connected and automated vehicles (CAVs) in conjunction with vehicle-to-everything (V2X) communication technologies offer a promising framework for advancing eco-driving methodologies \cite{shao_vtm}. By leveraging sophisticated communication systems to integrate real-time traffic data such as the speed, acceleration, and position information of preceding vehicles into advanced vehicle control algorithms, these technologies enable the optimization of vehicle speed profiles and powertrain outputs for energy savings. This not only lowers operational costs for drivers and businesses but also minimizes environmental impact by decreasing greenhouse gas emissions. 

In the realm eco-driving of CAVs , earlier research typically optimizes only vehicle speed, such as eco-cooperative adaptive cruise control (Eco-CACC) based on real-time sharing of traffic information in a platoon of vehicles \cite{7995884, shao2017robust}, eco-driving on a rolling horizon \cite{wang2014rolling, hu2016integrated}, and eco-driving at signalized intersections \cite{zhang2024eco, shao2018optimal}. Beyond speed optimization, powertrain operations can be further optimized based on vehicle demand and speed \cite{shao_vtm}. For instance, an integrated optimization framework for speed and powertrain management in hybrid electric vehicles (HEVs) was proposed in \cite{hu2016integrated} across various terrains. The powertrain management control determines the optimal power-split ratio between the engine and the battery for the HEVs.

In the context of battery electric vehicles (BEVs), one of the main challenges is range anxiety. Eco-driving for BEVs has the potential to improve overall operational efficiency and extend the range. For BEVs, optimal powertrain management strategies can be designed in addition to vehicle speed optimization. Recently, dual-motor powertrains have attracted considerable attention (e.g., Tesla Model S P100D). Due to considerations of power demand and cost, dual-motor BEVs typically do not utilize identical motor drives. The independent operation and distinct electrical performances of the two motors necessitate real-time determination and management of power distribution to ensure optimal performance, maximize energy efficiency, and maintain vehicle stability under varying driving conditions \cite{guo2020systematic,yang2021energy}. Current dual-motor EVs often rely on heuristic strategies for power splitting, which cannot always maintain system operations in the optimal regions or adapt to varying traffic conditions effectively. The integration of V2X information offers new opportunities to predict future driving conditions and optimally determine power distribution between the motors.

In order to optimize powertrain efficiency, few studies have developed optimization algorithms for multiple motors. With the advancement of in-vehicle computational resources, most existing work on eco-driving adopts model-based control design to solve the optimization problem in real-time. For example, in \cite{hu2018optimal}, the convex programming multi-power-source integration problem is implemented, and nonlinear model predictive control (MPC) based powertrain management strategies are proposed in \cite{guo2020systematic,yang2021energy}. Recently, model-free methods like reinforcement learning (RL) have gained traction due to their learning and adaptive capabilities, with numerous studies developing control algorithms based on this approach \cite{shi2021connected,9147717}. However, the RL approach requires offline training and faces challenges in robustness and the ability to handle hard constraints to ensure safety.

In this work, we present a novel real-time capable co-optimization algorithm that simultaneously optimizes vehicle speed and torque allocation between dual motors for connected and automated electric vehicles to achieve energy savings. The core of our approach lies in an optimal control algorithm, developed based on a control-oriented vehicle dynamics and powertrain model, specifically designed to facilitate real-time computation. To validate the effectiveness of the proposed algorithm, we conducted extensive simulations using traffic scenarios calibrated with real-world data from a signalized corridor in Chattanooga, TN. Our evaluation demonstrates that the controller not only enhances energy efficiency by up to 24.52\% compared to baseline strategies, but also maintains robust performance under various traffic prediction uncertainties. This robustness ensures reliable operation in dynamic and complex traffic environments. Additionally, the study highlights the algorithm's capability to manage power distribution between the two motors effectively, thereby optimizing overall vehicle performance and reducing operational costs.

% \begin{figure}[htbp]
%   \centering
%   \includesvg[width=0.4\textwidth]{fig/OPveh_11084_Tf}
%   \caption{test}
% \end{figure}

\section{ Electric Powertrain Modeling }
In this section, we formulated control-oriented vehicle dynamics and powertrain model, which is specifically designed to facilitate real-time computation without losing accuracy.

% \begin{equation}
%     P_{\text {req }}(t)=m \cdot a(t) \cdot v(t)+\frac{1}{2} m \cdot v(t)^{2} \cdot \frac{d v(t)}{d t}+m \cdot g \cdot \sin (\phi(t)) \cdot v(t)
% \end{equation}
% \subsection{Regenerative braking constrains}
% Regenerative braking is inefficient at low speeds, assume the cut off speed: $v_{\text{cutoff}}$

% Also, the regenerative braking torque can 

\subsection{Powertrain Model}
% \subsubsection{Vehicle Dynamics Model}
The longitudinal dynamic model is describe by \cite{he2023real}:

\begin{equation}
    \begin{aligned}
        T_m(k) n &= m a(k) + mg \sin(\phi(k)) \\
        &\quad + \mu_r mg \cos(\phi(k)) + \frac{1}{2} k_w v(k)^2 
        - F_b(k) \\
        &= m \cdot a(k) + F_g(k) + F_r(k) + F_a(k) - F_b(k) \\
    \end{aligned}
    \label{eq:1}
\end{equation}
\begin{equation}
     k_w =  C_D \rho_a  A
\end{equation}
where $m$ is the vehicle total mass, $a$ is vehicle longitudinal acceleration, $g$ is the acceleration due to gravity, $v$ is the vehicle speed; $n$ is the lumped drivetrain ratio which equals to the final drive ratio divided by effective tire radius, $F_t$ is the vehicle total traction force, $F_b$ is brake force,  $\phi$
is slope of roadway, $\mu_r$ is the rolling resistance coefficient. and $C_D,
    \rho_a, A $ are the drag coefficient, air density and frontal area contribute
to the coefficient of wind resistance $k_w$. $T_m$ represents the total motor torque and the proposed optimal control will determine the dual-motor power split into front motor torque $T_f$ and rear motor torque $T_r$:
\begin{equation}
    \begin{aligned}
        T_m(k)  & = T_f(k) + T_r(k)    
    \end{aligned}
\end{equation}
Relationship between motor speed  $\omega$ and vehicle speed is:
\begin{equation}
   \omega(k) = n \cdot v(k)
\end{equation}

\subsection{Vehicle Power Consumption Model}
Total power output of vehicle includes two parts~\cite{
bauer2014thermal}: 
\begin{equation}
    I_{\text{bat}}(k)U_{\text{bat}} = P_{\text{trac}}(k) + P_{aux}(k)
\end{equation}
where $P_{\text{aux}}$ is auxiliary power (e.g., infotainment, A/C system, battery cooling), $P_{\text{trac}}$ is traction power demand.
% $P_{\text{hvch}}$ is the power of high voltage cooling heater, $ P_{hvac}$ power of heating ventilation and air conditioner.

For simplicity, this paper focus on $P_{\text{trac}}$ without considering $P_{aux}$ optimization. The demand of traction power can be expressed as:
% \begin{equation}
%     P_{\text{trac}}(k) = 
%     \begin{cases} 
%      \left(v(k)\cdot (F_{ext}(k)+ m\cdot a(k)) \right) / \eta(\cdot)  &  \text{propelling} \\
%      \left(v(k)\cdot (F_{ext}(k) + m\cdot a(k)) \right) \cdot   \eta(\cdot),     & \text{braking.}
%     \end{cases}
%     \end{equation}

    \begin{equation}
        \begin{aligned}
            P_\text{\text{trac}}(k) = &\underbrace{T_{r}(k)\cdot v(k)\cdot n}_{P_{m1}} \cdot \eta_{f}(k) (n \cdot v(k), T_{r}(k) + \\
            &\underbrace{T_{f}\cdot v(k)\cdot n}_{P_{m2}} \cdot \eta_{r} (n \cdot v(k), T_{f}(k)) 
        \end{aligned}
    \end{equation}    
where $\eta_f(\cdot),\eta_r(\cdot)$ describes the power conversion efficiency between the battery and the motor. The efficiencies are modeled using consumption efficiency maps for the front and rear motor respectively~\cite{he2023real}.  

 % Typically, the operating points and corresponding power consumption of a motor can be obtained through testing or OEM data. 
By convention, $\eta$ is always less than 1 and follows:
 \begin{equation}
\eta(\omega,T) = \begin{cases}
\displaystyle \frac{\omega T }{P_{\text{map}}(\omega,T)} \times 100 \% & \text{for propelling}, \\
\displaystyle \frac{P_{\text{map}}(\omega,T)}{\omega T} \times 100 \%  & \text{for braking}.
\end{cases}
 \end{equation}
 
 Such piecewise relationship introduces complexity for optimization \cite{shao2019optimal}. We applied five-order polynomials to approximate the efficiency maps with one polynomial fitting for both propelling and regenerative braking regions.

\subsection{Battery Model}

The equivalent resistance battery model is used \cite{guo2020systematic}:
\begin{equation}
    I_{\text {bat }}(k)=\frac{U_{o c}-\sqrt{U_{o c}^2-4 P_{\text {bat }}(k) R_b}}{2 R_b}
\end{equation}
where \( P_{\text{\text{bat}}} \) represents the battery power, \( I_{\text{\text{bat}}} \) is the battery current; \( U_{\text{oc}} \) and \( R_b \) are the battery open-circuit voltage and internal resistance, these variables depends on temperature and SOC, and will get updated from powertrain every MPC update horizon. The SOC can be calculate as below:
% % \begin{equation}
% %     U = f_{U}(SOC(k), T_{\text{\text{bat}}})
% % \end{equation}
% where the $f_{U}$ is the function of temperature.

% where the $f_{U}$ is the function of temperature.
\begin{equation}
    SOC(k+ 1) = SOC(k) - \frac{I_{\text{\text{bat}}}(k) \Delta t}{C_{\text{\text{bat}}}}
\end{equation}
where $\Delta t$ is time step, \( C_{\text{\text{bat}}} \) is the battery nominal capacity.

\section{Optimal Control Problem Formulation}
The optimal control is implemented in a MPC framework \cite{shao_vtm}  that updates optimal control commands once new traffic information is available.  The proposed control framework is depicted in a schematic diagram as illustrated in \cref{fig:control_diagram}. Our algorithm processes predictive traffic information data obtained through the communication module at each control step, alongside real-time feedback of vehicle drivetrain information, to generate both the optimal vehicle speed and power split for the dual motors via the motor controller. Since traffic conditions are highly dynamic and varying, we aim to have a control update horizon of less than 1 second. To ensure such vehicle speed updates and communication of information can occur in real time, the optimization problem must be sufficiently simple to solve efficiently but without sacrificing the accuracies significantly. Therefore, the following assumptions are made:

\begin{figure}[h]
    \centering
    \includegraphics[width=0.48\textwidth]{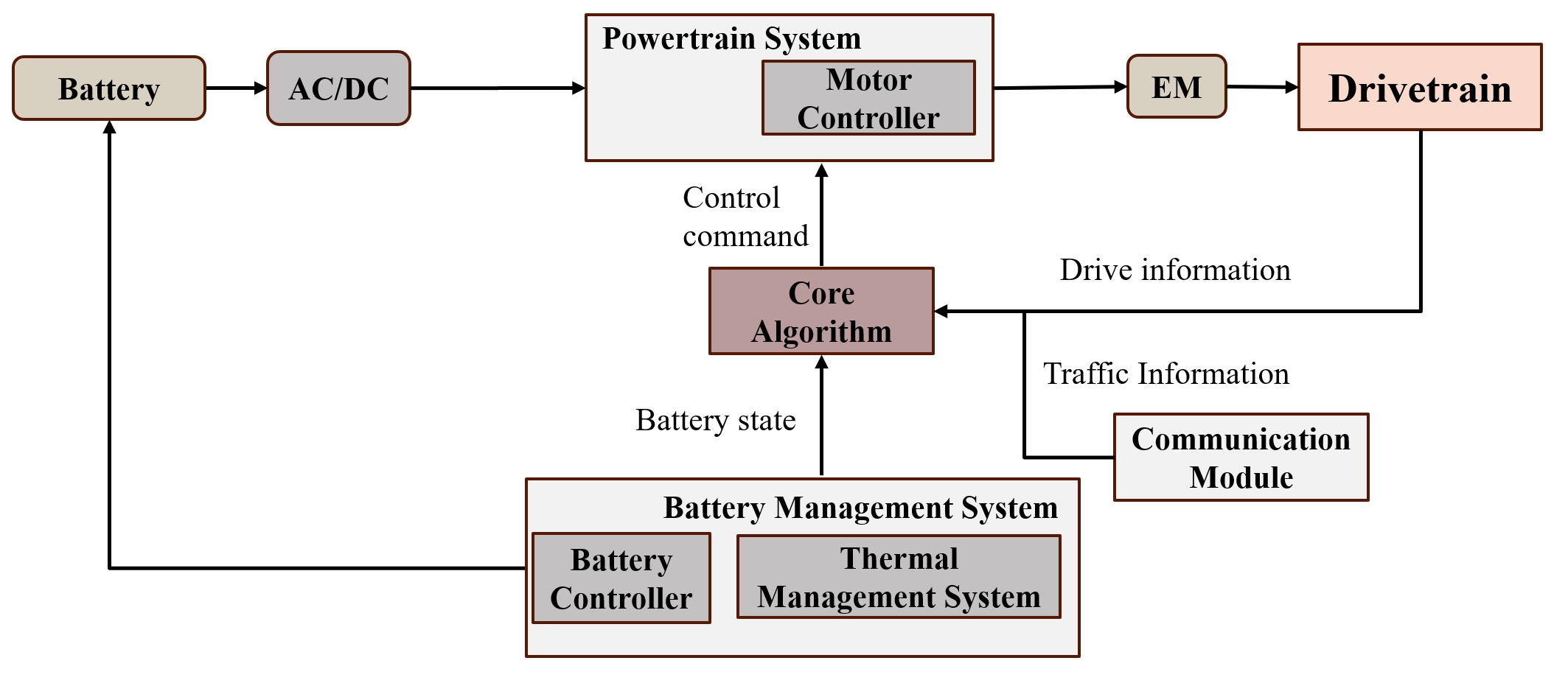}
    \caption{Control diagram of the proposed framework}
    \label{fig:control_diagram}
\end{figure}

\begin{asm}[Battery performance]
Battery aging effects are negligible over the short time horizon considered. It is assumed that the battery capacity remains constant. 
\end{asm}
\begin{asm}[Motor speed consistency]
    \label{asm:motor}
    Dual Motor rotate at the same speed. The dual motors operate at a uniform speed. Tire slips are neglected, thereby maintaining a linear correlation between motor speed and vehicle speed.
\end{asm}
\begin{asm}[Car following behavior]
Ego vehicle always follow the same preceding vehicle throughout the traffic corridor, see \cref{section:results}.
\end{asm}
\begin{asm}[Thermal management system]
There is a separate battery management system (BMS) that controls the temperature of the battery, motor and other components. Every MPC update step, the BMS will send the current battery temperature and status to the optimal controller, as illustrated in \cref{fig:control_diagram}. The proposed controller assumes that temperature holds constant during each MPC horizon. 
\end{asm}

\subsection{State-space Model}
\label{subsection:statespacemodel}
The overall discrete state space model is as the following.

\begin{equation}
\begin{aligned}
    \mathbf{X}(k+1) &=
        \begin{bmatrix}
            d(k+1) \\
            v(k+1) \\
            %T_{\text{bat}}(t),
            SOC(k+1) 
        \end{bmatrix} \\&
        =
        \begin{bmatrix}
            d(k)+ \Delta t \cdot  v(k)\\
            v(k) +\Delta t \cdot a(k)    \\
            SOC(k) - \Delta t\cdot (\frac{U_{o c} - \sqrt{U_{o c}^2-4 P_{\text{bat}}(k) R_{\text{bat}}}}{2 R_{\text{bat}} C_{\text{bat}}}) \\
        \end{bmatrix}
    \\
\end{aligned}
\end{equation}
\begin{align}
    \mathbf{u}(k) &= [
        T_f(k),
        T_r(k),
        P_{\text{bat}}(k),
        F_{\text{b}}(k)    
        % T_{m_r}(t), T_{m_f}(t) 
    %, P_{\text{trac}}(t)
    % , P_{hvac}(t) 
    ]^T \\
    \mathbf{W}(k) &= [\hat{d}_{p}(k), 
    \phi(k)
    %  ,T_{amb},
    % P_{\text{hvch}}
    % ,P_{\text{hvac}}
    ]^T
\end{align}
where $\mathbf{u}(k)$ is system control input, $\mathbf{W}(k)$ is system external input which includes the predicted preceding vehicle location $\hat{d}_p$; $a$ is can be computed based on \eqref{eq:1}, $d$ is the position of the ego vehicle.

\subsection{Constraints}
\label{subsection:constraints}
\subsubsection{Traffic Dynamics Constraints}
\paragraph{Car Following Constrains}

The robust and recursively feasible car-following constraints are defined as \cite{shao2020eco}:
\begin{equation} \begin{aligned} & d(k) \geq \hat{d}_p(k)
-d_{\max }-s_1(k) \\ & d(k) \leq \hat{d}_p(k)
-\left(d_{\min }+h_{\min} \cdot v(k)\right)+s_2(k) \end{aligned}
\label{eq:soft}
\end{equation}
where $h_{\min}$ is the safety time headway, $d_{max},d_{min}$ are the maximum following distance and minimum following distance.  $s_1,s_2$ are slack variables to enforce the constraints as soft constraints to improve feasibility of MPC. Slack variables must be positive and are penalized in cost function \cref{subsection:costfunction}.
\paragraph{Signal Constraint}
Ego vehicle should pass the intersection only if the signal is green \cite{shao2020eco}, 
\begin{equation}
    d(t_g) \leq d_{sig} \leq d(t_r)
\end{equation}
$t_r, t_g$ time instances of the beginning of next red light and green light, $d_{sig}$ is location of next intersection .
\paragraph{Speed Limit Constraints}
Speed  should limited by the current corridor during the driving cycle: 
    \begin{equation}
    0 \leq v(k) \leq v_{\max}
    \end{equation}
\subsubsection{Vehicle Dynamics Constraints}
\paragraph{Motor Speed Constraint}
The motor speed  should be no less than zero and no more than the maximum value:

\begin{equation}
    \begin{aligned}
         0 \leq n \cdot v(k)& \leq \omega_{m_{max}}
    \end{aligned} 
\end{equation}

\paragraph{SOC Constraints}
The battery state of charge (SOC) should be within the range of 0 and 1:
\begin{equation}
    0 \leq SOC(k) \leq 1
\end{equation}

\paragraph{Brake Force Constraints}
The frictional brake force should be positive and limited by maximum brake force:
\begin{equation}
    0 \leq F_b(k) \leq F_{b_{\max }}
\end{equation}
% \paragraph{Regenerative Brake Constraints}
% In general, regenerative brake will disable in very high speed 

\subsubsection{Powertrain Constraints}

\paragraph{Side Slip Constraints}
\cref{asm:motor} required the direction of rotation of the front and rear motors to be the same:
\begin{equation}
    T_f(k) \cdot T_r(k) \geq 0
\end{equation}

\paragraph{Jerk Constraints}
In order to ensure the comfort for the passengers, the jerk should be limited and it can implement by constraint the torque change rate:
\vspace{-2mm}
\begin{equation}
    \begin{aligned}
         & -\Delta T_{f_{\max}} \leq T_{f}(k+1)-T_{f}(k) \leq \Delta T_{f_{\max}}  \\
         & -\Delta T_{r_{\max}} \leq T_{r}(k+1)-T_{r}(k) \leq \Delta T_{r_{\max}}  
    \end{aligned}
\end{equation}
$\Delta T_{f_{\max}},\Delta T_{r_{\max}}$ is determined by maximum comfortable jerk value $j_{\max}~\cite{he2023real}$.

\paragraph{Max Torque Constraints}

The maximum and minimum motor torque should follow the physical limitations (see \cref{fig:motor_combined}):
\vspace{-1.5mm}
\begin{equation}
    \begin{aligned}
     & -T_{f_{max}}(n \cdot v(k))  \leq T_{f}(k) \leq T_{f_{max}}(n \cdot v(k)) \\
     & -T_{r_{max}}(n \cdot v(k)) \leq T_{r}(k) \leq T_{r_{max}}(n \cdot v(k)) 
    \end{aligned}
\end{equation}

%%%%%%  Thermal Management System 
% \subsubsection{Thermal Management System}

% High-voltage coolant heater (HVAC) and high-voltage coolant chiller(HVCH) are used to control the battery temperature. The power of the HVAC and HVCH can be expressed as:
% \begin{align}
%     & 0 \leq P_{hvch} \leq P_{hvch,max} - P^c_{hvch}(T_{amb}) \\
%     & 0 \leq P_{hvac} \leq P_{hvac,max}
% \end{align}

% And battery temperature should be within the range of 0 and 60 degree Celsius:
% \begin{equation}
%     T_{bat,min}(k) \leq T_{\text{bat}}(k) \leq T_{bat,max}(k)
% \end{equation}

% \subsection{Discretization}
% The model proposed in this paper is based on continuous-time, in optimization
% discrete time model is required and it can be converted to discrete time model
% by using Euler method.

% \begin{equation}
%     \boldsymbol{x}(k+1)=\left[\begin{array}{l}
%             x_1(k+1) \\
%             x_2(k+1)
%         \end{array}\right]=\left[\begin{array}{l}
%             d(k)+d t \cdot v(k) \\
%             v(k)+d t \cdot a(k)
%         \end{array}\right]
% \end{equation}
% where $x_1$ represents the distance from the initial points and $x_2$ represents the vehicle speed with units (m/s).
% \paragraph{Initial Conditions} 
% Assumed that initial vehicle location and speed are known
% \begin{equation}
%     d(1) = d_0, v(1) = v_0
% \end{equation}

\subsection{Cost Function Formulation}
\label{subsection:costfunction}
The general optimization problem is to minimize a cost function $J$:

\begin{equation}
\begin{aligned}
    &\min \ J(\textbf{X}, \mathbf{W}) \\  
    & \text{s.t.} \ g(\textbf{X},\mathbf{W}) \leq 0, i = 1,\dots,n_c
\end{aligned}
\end{equation}
where $g(\textbf{X},\mathbf{W})$ denotes generalized constraints which includes system equation \cref{subsection:statespacemodel} and all constraints from \cref{subsection:constraints}, $n_c$ denotes the number of constraints. And objective function  $J(\textbf{X},\mathbf{W}) =J_p+ J_f + J_t $ can be divided into three terms, first term $J_p$ penalize the over acceleration and battery power output.  Second term represents terminal constraints which try to follow up the speed and distance of the proceeding vehicle. The third term is the penalize of slack variables to implement the soft car following constraints \eqref{eq:soft}. These cost functions are formulated as:
% \begin{equation}
%     J_p = \int_{t_0}^{t_f}  w_1 a^2 (t) + w_2 u^2(t) dt
% \end{equation}

\begin{equation}
    J_p = \sum_{k=1}^{N} w_1 a^2 (k) + w_2 P_{bat}(k)
\end{equation}
\begin{equation}
    \begin{aligned}
        J_{t} = &w_3 (\hat{d}_{p}(N)- d(N) - h_{\text{head}}v_p(N) - d_{min})^2  \\
        &+ w_4 (v(N) - \hat{v}_{\text{p}}(N))^2
    \end{aligned}
\end{equation}

\begin{equation}
    J_{f} = \sum_{k=1}^{N}  (w_5 s_1^2(k) + w_6 s_2^2(k))
\end{equation}
where $h_{\text{head}}$ represents the general time headway, $N$ represents the length of prediction horizon, $w_i$ are weights whose values are as in \cref{tab:weight}.

% In modeling, we want the followed car within the Safety zone:
% \begin{equation}
%     \begin{aligned}
%         J & =\frac{1}{2} \int_{t_0}^{t_f} w_1(V_p(t) - V(t))^2 + w_2 a^2 (t) + w_3(V_e(t) -
%         P_{l}(V_e(t),SOC)) dt                                                               \\
%           & =  (X - X_{ref})^T Q (X - X_{ref}) + U^T R U + (V_e - P_l(V_e))^2
%     \end{aligned}
% \end{equation}
% where
% \begin{equation}
    % X_{ref} = \left[ \begin{array}{c} P_p + d_{safe} \\ V_p \end{array} \right]
% \end{equation}

% \subsection{Three Layer Programming}
% Since we assume that traffic model can only provide short range of prediction, total optimization horizon is based on the traffic prediction horizon. In order to maintain the continuity, we decide to use three layer programming to solve the problem.

% \subsubsection{First Layer: Traffic Prediction Horizon }

% \begin{equation}
%     \min_{u^*} J(\cdot) 
% \end{equation}

% \subsubsection{Second Layer: Temperature Control and Powertrain Horizon}

\section{ Results}
\label{section:results}

\subsection{Simulation Setup}

The discretized optimal control optimal control problem is solved in MATLAB using CasADi~\cite{andersson2019casadi} with solver IPOPT \cite{wachter2006implementation}. The simulation is run on a desktop PC with Intel\texttrademark 14900K CPU at 3.20 GHz and 64 GB RAM.
Prediction horizon for MPC layer is set as $15s$, and the control update interval is set as $1s$. In order to trade off solver time and accuracy, the discredited time $d_t$ is set as $0.1s$. The solver IPOPT can solve the optimization problem within $1s$ which is smaller than the control update interval and indicates the real-time implementation capability, \cref{tab:parameter} shows the list of vehicle parameter used in the model.

\begin{figure}[tb]
    \centering
    \includegraphics[width=0.48\textwidth]{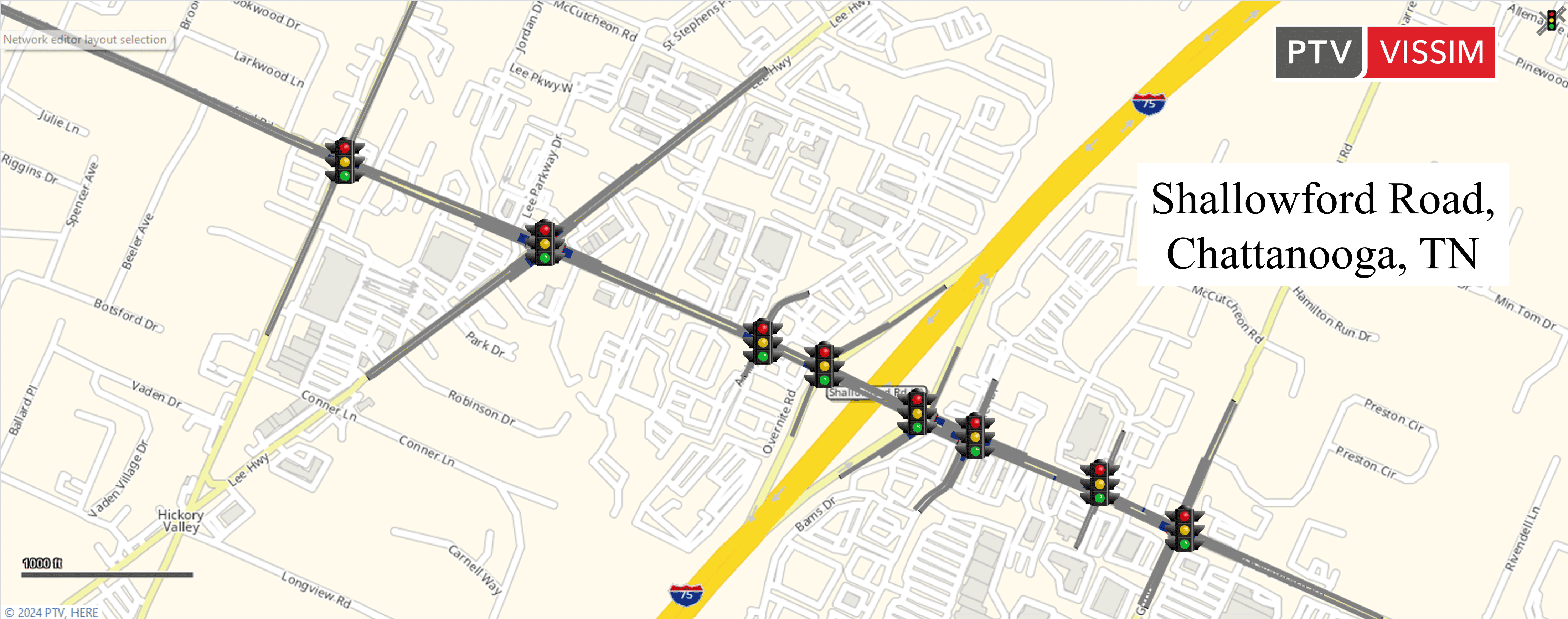}
    \caption{Traffic simulation scenario in VISSIM.}
    \label{fig:vissim}
\end{figure}

The simulation scenario is developed using VISSIM, a widely recognized state-of-the-art traffic micro-simulation software. \cref{fig:vissim} shows the VISSIM simulation network of the Shallowford Road corridor in Chattanooga, Tennessee. The scenario is built and calibrated with real-world data, which includes detailed traffic volumes, speed information, and traffic signal controller configurations. The simulation scenario has been validated \cite{yuan_ecodriving} to accurately represent real-world traffic characteristics, such as traffic volume, speed, and travel time. To evaluate the proposed optimal control strategy, various vehicles were randomly selected from the VISSIM simulation as preceding vehicles.

\begin{table}[h]
    \centering
    \vspace{-3mm}
    \caption{List of vehicle basic parameters}
    \footnotesize  % 调整字体大小
    \setlength{\tabcolsep}{3pt}  % 调整列间距
    \begin{tabular}{p{3cm} p{1.5cm} p{2cm}}
        \toprule
        \textbf{Parameter} & \textbf{Notation} & \textbf{Value} \\
        \midrule
        \multicolumn{3}{c}{Vehicle} \\
        \midrule
        Vehicle mass & $m$ & $1780\ kg$ \\
        Battery capacity & $C_{\text{bat}}$ & $150\ Ah$ \\
        Drag coefficient & $C_D$ & $0.306\ kg/m$ \\
        Air density & $\rho_a$ & $1.205\ kg/m^3$ \\
        Frontal area & $A$ & $2.200\ m^2$ \\
        Wind resistance coeff. & $k_w$ & $0.881$ \\
        Rolling resistance & $\mu_r$ & $22.910$ \\
        Gravity constant & $g$ & $9.81\ m/s^2$ \\
        Lumped ratio & $n$ & $22.910$ \\
        Battery resistance &  $R_{bat}$ & $0.228 \Omega$ \\
        Max motor speed & $\omega_{\max}$ & $ 1400 \  rad/s$ \\
        
        \midrule
        \multicolumn{3}{c}{Constraint} \\ 
        \midrule
        Max following distance & $d_{\text{max}}$ & $80\ m$ \\
        Min following distance & $d_{\text{min}}$ & $1\ m$ \\
        Max acceleration & $a_{\text{max}}$ & $3\ m/s^2$ \\
        Min acceleration & $a_{\text{min}}$ & $-3\ m/s^2$ \\
        Time headway & $h_{\text{head}}$ & $2.5\ s$ \\
        Time safety & $h_{\min}$ & $0.5\ s$ \\
        Initial position & $d_{p_0} - d_0$ & $40\ m$ \\
        % Regen. cutoff speed & $v_{\text{cutoff}}$ & $2\ m/s$ \\
        Max brake force & $F_{b_{max}}$ & 15000 N   \\
        Initial state of charge & $SOC_0$ & $0.8$ \\
        Speed limit & $v_{\text{max}}$ & $20\ m/s$ \\
        Max Jerk    & $j_{max}$ & $3.0\  m/s^3$ \\ 
        Torque change rate & $\Delta T_{m_{\text{max}}}$ & $150\ Nm/s$ \\
        \bottomrule
    \end{tabular}
    \label{tab:parameter}
\end{table}
\vspace{-2mm}
\begin{table}[h]
    \centering
    \caption{Cost function weights}
    \begin{tabular}{lllllll}
        \toprule
        $w_1$ & $w_2$ & $w_3$ & $w_4$ & $w_5$ & $w_6$ \\
        \midrule
        $10^{1.5}$ & $10^{-3}$ & $10$ & $10^2$ & $1$ & $1$ \\
        \bottomrule
    \end{tabular}
    \label{tab:weight}
    \vspace{-3mm}
\end{table}

\subsection{Evaluation}

To evaluate the algorithm, we benchmark the performance of the optimal control strategy with the performance of the preceding vehicle. The initial distance to the preceding vehicle is set at $40m$, and the initial speed is the same as that of the preceding vehicle. We assumes that the preceding vehicle shares the same powertrain model and vehicle parameters for fair comparison. We evaluate the performance of the proposed method by comparing the final battery SOC with that of the preceding vehicle, using the relative improvement in SOC metric \( R_{\text{SOC}} \):
\begin{equation}
    R_{\text{SOC}} = \frac{SOC_{e} - SOC_{p}}{SOC_{0} - SOC_{p}} \times 100\%
\end{equation}
where \( SOC_{e} \), \( SOC_{p} \) is the state of charge of the ego vehicle and preceding vehicle after driving cycle, \( SOC_{0} \) is the initial $SOC$ of the ego vehicle.

 Following typical commercial dual motor EVs, the front motor is designated as an Induction Motor (IM), while the rear motor is identified as a Permanent-Magnet Synchronous Motor (PMSM). The IM exhibits greater torque at lower speeds, whereas the PMSM demonstrates enhanced efficiency at lower speeds. All motor data are obtained based on actual motor data in \cite{he2023real,yang2021energy}. The use of different motors allows for a more optimal allocation of torque compared to the traditional rule-based method, necessitating a quantitative comparison with the proposed method. We conducted ablation studies that eliminate the influence of optimized speed. We then apply a rule-based method to allocate the required torque and compare it against the original allocation from co-optimization. The relative improvement in power consumption, \( R_{m} \), is calculated as:

\begin{equation}
    R_{m} = \frac{P_{\text{trac,rule}} - P_{\text{trac,opt}}}{P_{\text{trac,rule}}} \times 100\%
\end{equation}
where \( P_{\text{trac,rule}} \) is the power consumption of the rule-based method, and \( P_{\text{trac,opt}} \) is the power consumption of the proposed method.

% The quantitative metric is defined as the relative improvement in state of charge, 

 We adopt a fixed ratio, rule-based power allocation strategy following~\cite{yang2021energy}. In this method, controller always divide torque demand $T_d$ with fixed ratio $N_f:N_r$ as following:
\begin{equation}
\begin{aligned}
    &T_f(k) = \frac{N_f}{N_f+N_r} T_d(k)\\
    &T_r(k) = \frac{N_r}{N_f+N_r} T_d(k)
\end{aligned}
\end{equation}
where $T_d(k)$ can be calculated with \eqref{eq:1} when speed is determined. When the distribution exceed the maximum torque, this rule-based strategy will allow extra torque allocated into another motor. In this Simulation,  ratio is established at $N_f = 1, N_r =1$. 
%Note that the preceding vehicle is also assumed to employ this strategy.

\begin{figure}[hbt]
    \centering
    \includegraphics[width=0.46\textwidth]{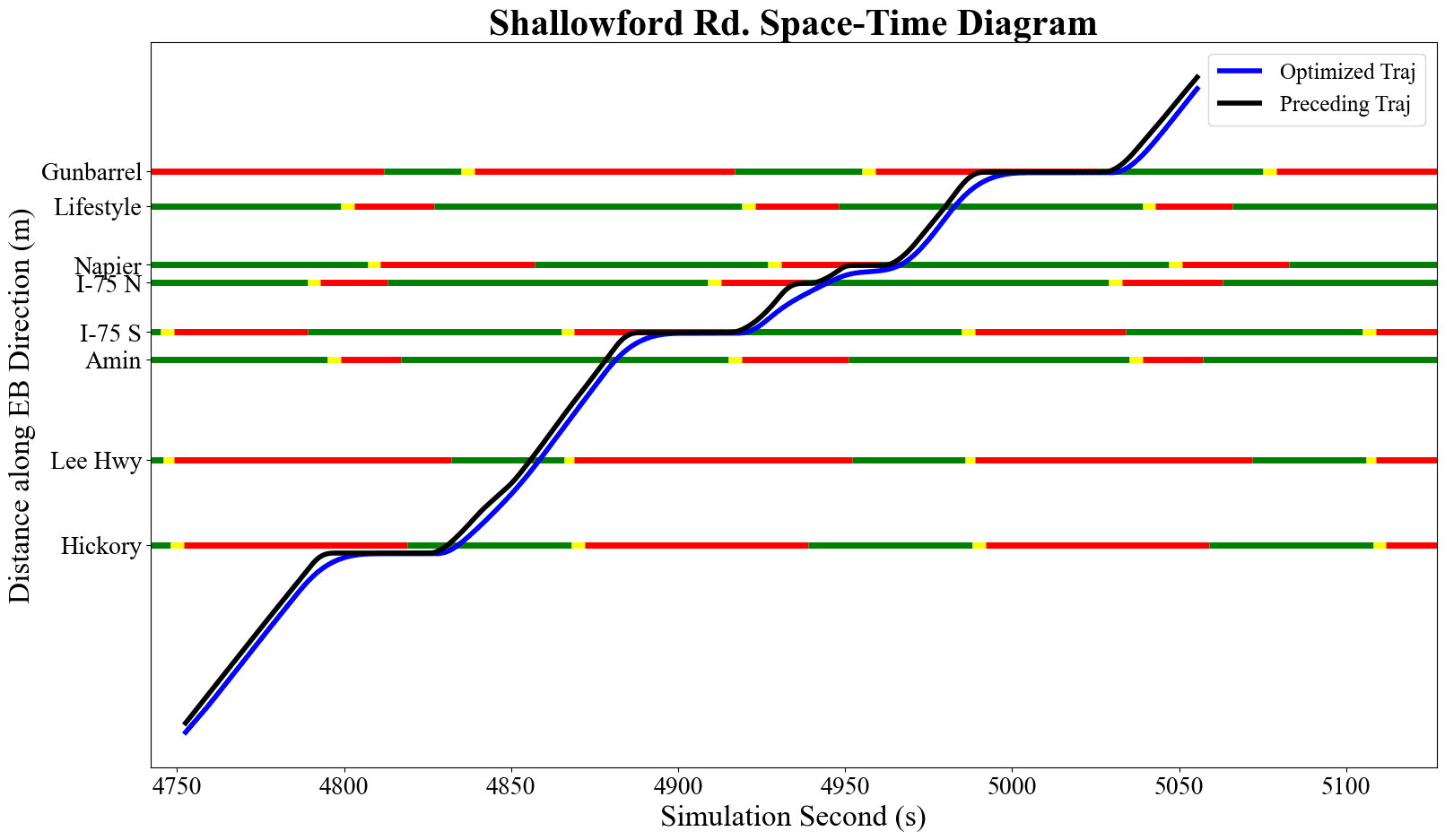}
    \caption{Optimized trajectory of the ego vehicle and preceding vehicle with traffic light status.}
    \label{fig:tlc}
\end{figure}

\paragraph{Ideal prediction}
Let we assume the prediction is ideal with no uncertainties first. \cref{fig:tlc} shows trajectory of the vehicle id 11084  with traffic light of the different intersections across this corridor. The optimized trajectory allows for the avoidance of lengthy periods of acceleration and deceleration, as well as the unnecessary expenditure of time spent at red lights.

\begin{figure}[htb]
    \centering
    \subfigure[State graph]{
        \includegraphics[width=0.43\textwidth]{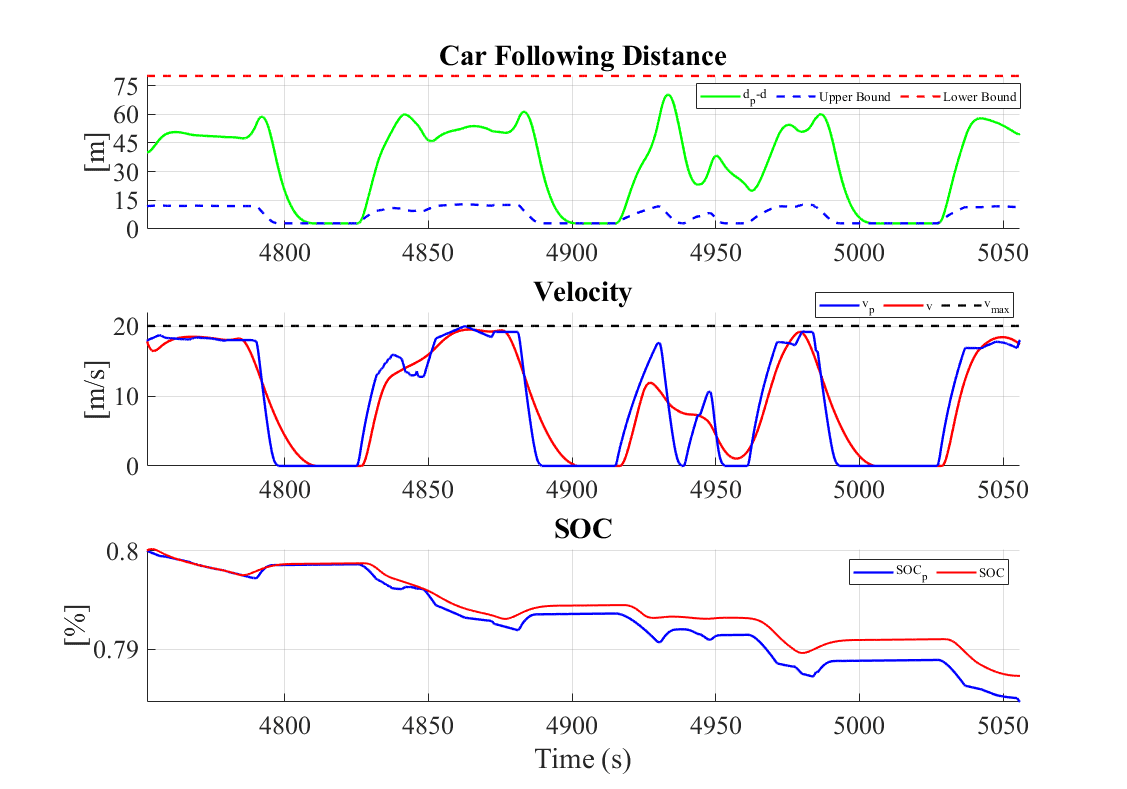}
        \label{fig:sim1a}
    }
    \hfill
    \vspace{-2mm}
    \subfigure[Powertrain graph]{%
        \includegraphics[width=0.43\textwidth]{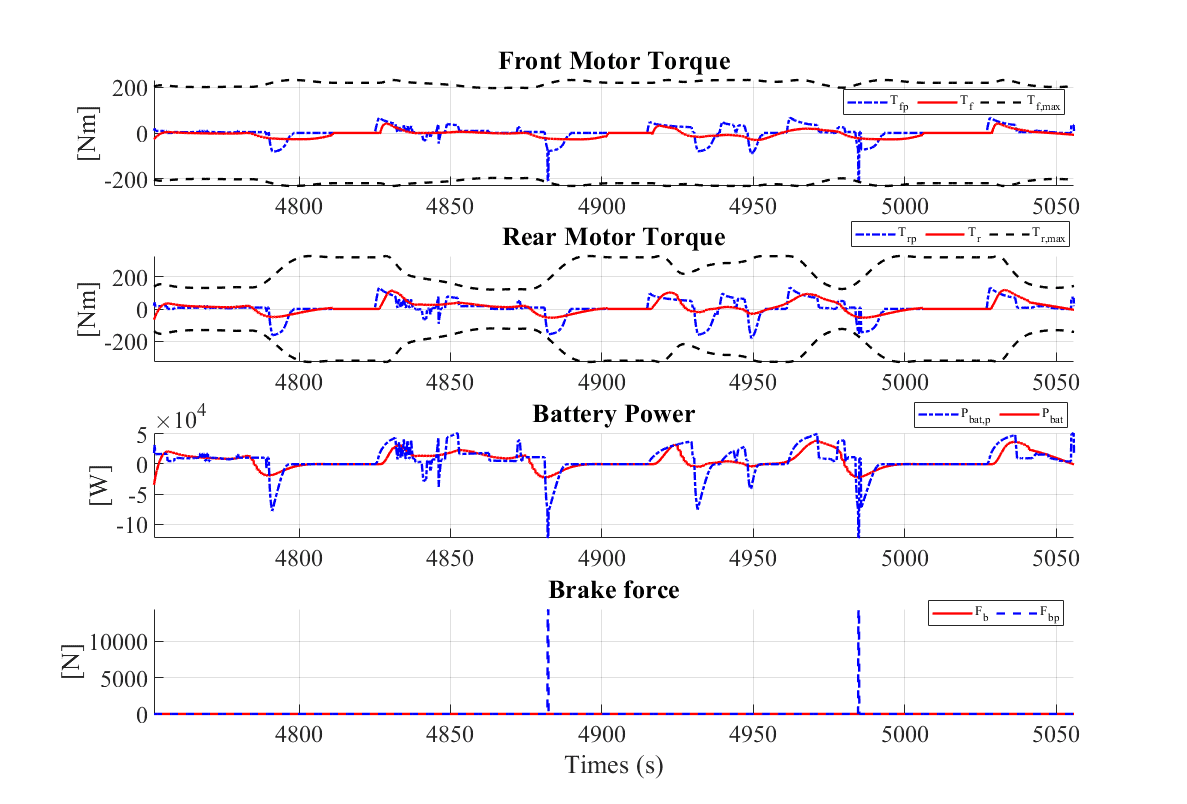}
        \label{fig:sim1b}
    }
    \caption{State and powertrain graphs with the preceding vehicle id 11084.}
    \vspace{-4mm}
    \label{fig:sim1_combined}
\end{figure}

\cref{fig:sim1a} shows the ego vehicle and preceding vehicle comparison with state variables. The ego vehicle is capable of regulating its speed in a manner that minimizes energy consumption when the predictive model is precise. This includes situations such as precise regenerative braking to avoid frequent acceleration and deceleration. Eventually it brings the $18.64\%$ power consumption reduction compared with the preceding vehicle. \cref{fig:sim1b} show the powertrain details about torque allocation and power consumption and brake force.  In comparison to the preceding vehicle, no brakes are employed, with both systems facilitating the motor's operation in regenerative braking scenarios.

To illustrate the generality of the proposed algorithm, we conducted simulations for different vehicles. Table \ref{tab:benefit} presents the results from these cases. As shown, the co-optimization achieved 12.80-24.52\% energy reduction while optimal dual-motor powertrain management alone achieved 3.16-4.40\% benefits.

\begin{table}[h]
    \centering
    \vspace{-2mm}
    \caption{SOC benefit and average solution time for horizon $15s$}
    \begin{tabular}{lcccc}
        \toprule
        Vehicle ID & $R_{soc}$  &$R_m$ & Avg Sol. time (s)&    Period (s) \\
                                                                   \midrule
         101  &  12.80\%    & 3.92\%  & 0.863     & 163.4 \\
         9799 &  24.52\%   & 4.40\%   & 0.917     & 293.7 \\
         10196 & 13.54\%    & 3.16\%   & 0.962     & 363.9 \\
         11084&  18.65\%    & 3.54\%  & 0.837     & 303.9 \\
        \bottomrule
    \end{tabular}
    \label{tab:benefit}
\end{table}

\begin{figure}[htb]
    \centering
    \subfigure[Front Motor]{%
        \includegraphics[width=0.4\textwidth]{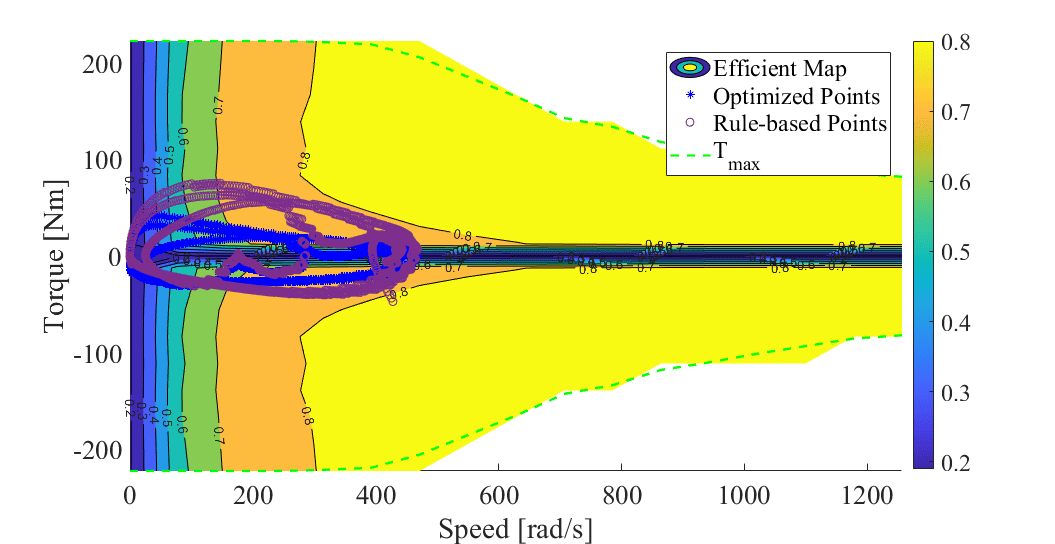}
        \label{fig:motorf}
    }
      % \vspace{-2mm}
    \hfill
    \subfigure[Rear Motor]{%
        \includegraphics[width=0.4\textwidth]{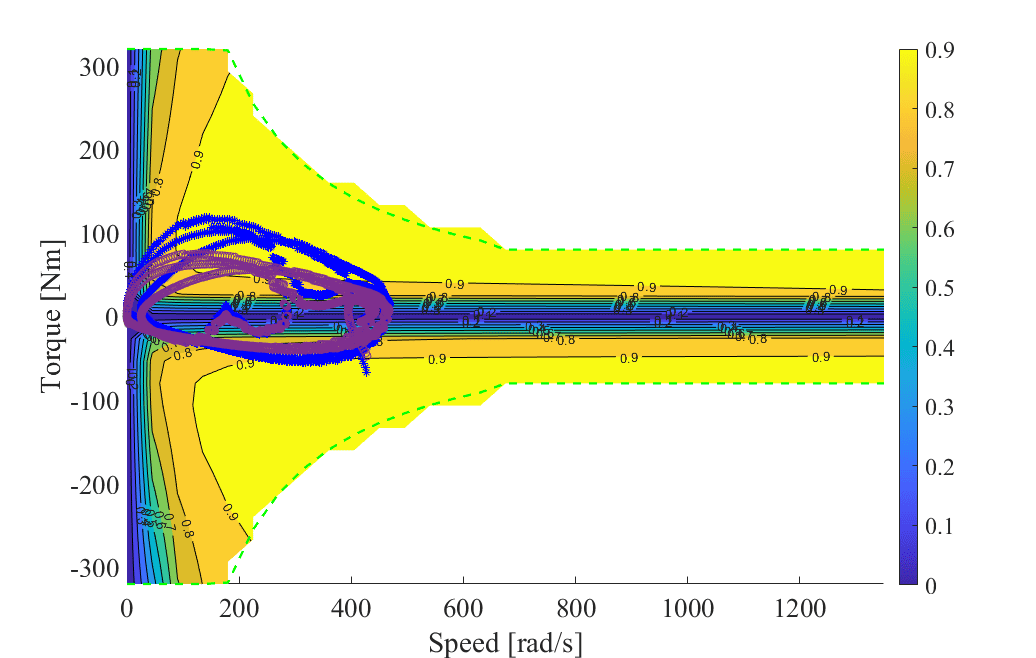}
        \label{fig:motorr}
    }
    \caption{Efficiency map and operating points for front and rear motors}
    \label{fig:motor_combined}
    \vspace{-3mm}
\end{figure}

% Motor map
\cref{fig:motorf} and \cref{fig:motorr} show the efficient map and operating points points for front and rear motors. As shown in the figure, the PMSM demonstrates higher efficiency than the IM within the rotational speed range of 100 to 500 rad/s. Consequently, the algorithm tends to allocate more torque to the rear motor. Compared to the rule-based distribution, this approach results in a greater number of operating points where the efficiency over 90\%.

\paragraph{Prediction with uncertainty}

\begin{figure}[hbt]
    \centering
    \includegraphics[width=0.48\textwidth]{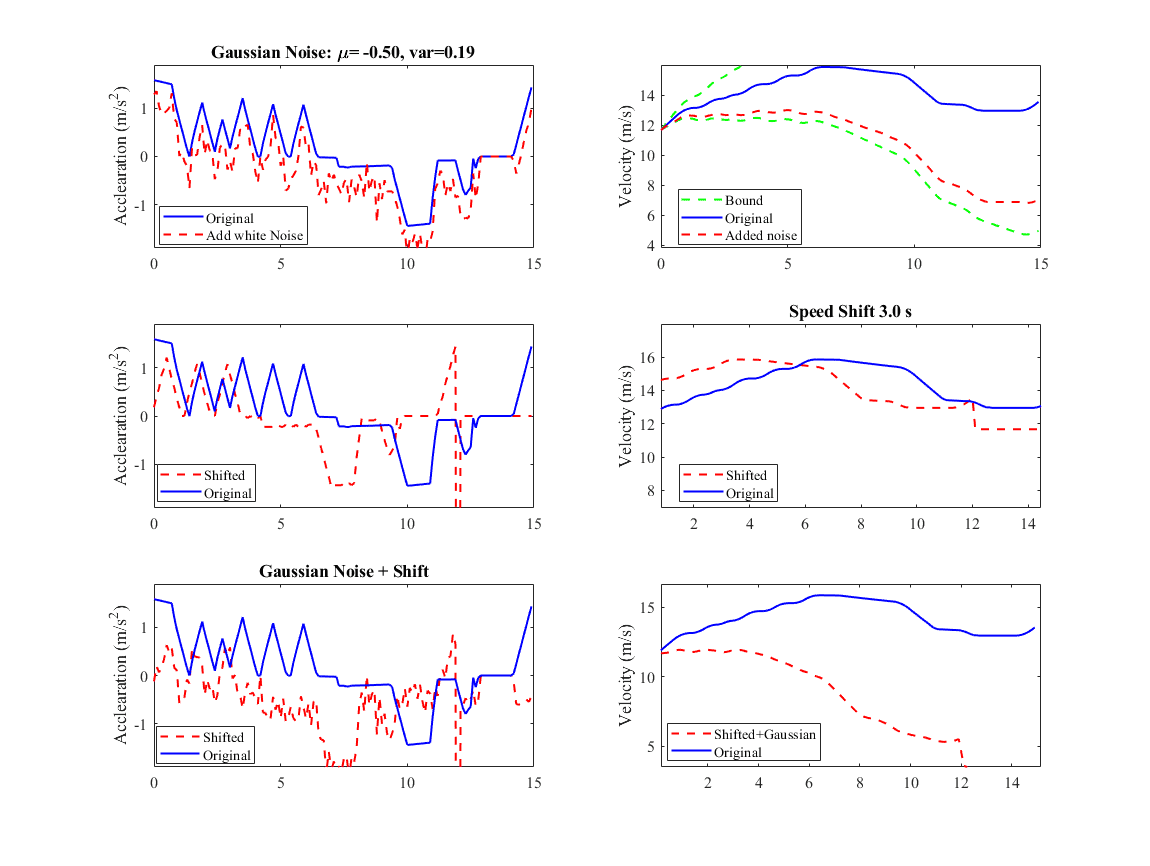}
    \vspace{-1mm}
    \caption{Predicted preceding vehicle trajectory with three types of noise}
    \label{fig:add_noise}
\end{figure}

Given that predictions cannot be perfectly accurate, it is essential to analyze scenarios with  prediction uncertainties. Small prediction errors can accumulate over time. Thus, we consider two types of uncertainties, as illustrated in \cref{fig:add_noise}. 

\textit{Time-increasing uncertainties:} The uncertainties of preceding vehicle's speed can be time-increasing. Based on the Wiener process~\cite{skorokhod1982studies}, Gaussian noise with variance $\sigma^2$ is added to the acceleration at each time step to simulate time-varying speed prediction uncertainties.

At each time step, the acceleration with noise is given by:
\begin{equation}
    \tilde{a}_p(i) = \hat{a}_p(i) + \mathcal{N}(0, \sigma^2),
\end{equation}
and the predicted speed is obtained by integrating the noisy acceleration:
\begin{equation}
\tilde{v}_p(k) = \sum_{i=0}^{k} \hat{a}_p(i) \Delta t = \hat{v}_p(k) + \mathcal{N}\left(0, \sigma^2 k (\Delta t)^2\right),
\end{equation}
where \(\tilde{v}_p(k)\) is predicted speed with Gaussian uncertainty and \(\mathcal{N}\left(0, \sigma^2 k (\Delta t)^2\right)\) represents the cumulative Gaussian noise effect on speed over \(k\) time steps.

\textit{Phase shift uncertainties:} The traffic prediction often requires prediction on timing of traffic signal light changes. Timing of actuated signal controllers depends on the real-time traffic demand measured from traffic detectors and are not fixed or known ahead of time~\cite{zhang2024eco}. The uncertainties in traffic signal light timing estimation could results in a phase shift effects on the predicted preceding vehicle. For example, the preceding vehicle might only accelerate 5 seconds later than the predicted acceleration time due to uncertainties in green light change timing. To simulate these effects, we introduce a time shift in the speed prediction, modeled as:
\begin{equation}
    \breve{v}_p(k) = \hat{v}_p(k + N_s),
\end{equation}
where \(\breve{v}_p(k)\) represent the predicted speed after applying the phase shift, and \(N_s\) is the number of time steps shifted, randomly selected from the range \(N_s \in [-P_s/2, P_s/2]\), with \(P_s\) representing the maximum shift magnitude. To ensure the data integrity, previous speed values are used to pad the shifted portion.

\begin{figure}[hbt]
    \centering
    \vspace{-2mm}
    \includegraphics[width=0.46\textwidth]{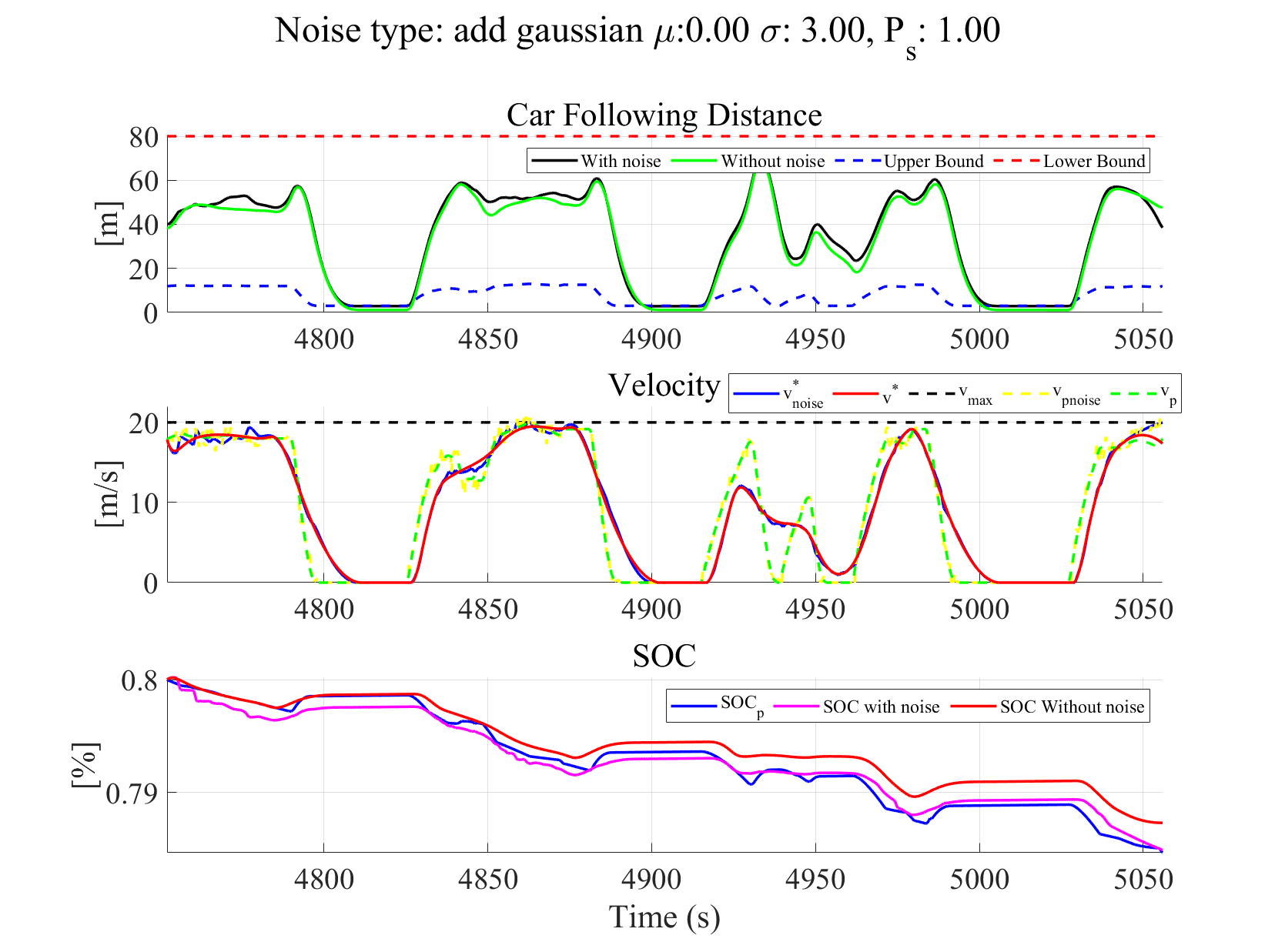}
    \caption{ State graph of noisy prediction results by adding Gaussian noise and Random time shifted.}
      \vspace{-3mm}
    \label{fig:noise_results}
\end{figure}

To more accurately reflect the inherent uncertainty associated with these phenomena, we employ a randomized time shift within the range \(P_s\) for each MPC update cycle. \cref{fig:add_noise} illustrate the prediction results of the preceding vehicle after adding Gaussian noise, time shift noise, and combined noise. It can be observed from figure that adding Gaussian noise to the acceleration results in predicted velocities with increasing errors over time, while the combination of the two uncertainties leads to significant changes in the predicted speed. \cref{fig:noise_results} displays the simulation results for vehicles under uncertainties and their benefits.  An analysis of the figure reveals that the introduction of substantial noise to the acceleration significantly increases the discrepancy between the predicted speed profile and the noiseless profile. Additionally, there is a notable decrease in the SOC benefit. \cref{tab:noise} summarizes the benefits obtained under different noise and shift conditions.  Controller is able to maintain benefit despite the added uncertainties in all scenarios, demonstrating robustness of the proposed algorithm under both types of uncertainties and continues to achieve energy savings.

% inkspacelatex=false,

\begin{table}
    \caption{Noisy prediction results for different noise type in \textit{veh 11084} }
    \centering
    \begin{tabular}{cccc}
        \toprule 
         Gaussian Adding Noise     &  $P_s (s)$ &  $R_{soc} (\%)$ \\
         \midrule
        $\sigma=0.20 \quad \mu=0.50 $     & 0.0s   &   17.40  \\
         
        $\sigma=0.20 \quad \mu=0.50 $     & 0.0s   &   10.83  \\
        $\sigma=0.20 \quad \mu=-0.50$    & 0.0s  &    23.58 \\
        $\sigma=0.75 \quad \mu=0.00 $     & 0.0s   &   18.24   \\
        $\sigma=3.00 \quad \mu=0.00$    & 0.0s  &    1.58\\
        $\sigma=0.00 \quad \mu=0.00 $ &     3.0s   &   16.92    \\
        $\sigma=0.00 \quad \mu=0.00 $ &     5.0s   &   14.59    \\

        $\sigma=0.50 \quad \mu=0.00 $ &    3.0s   &    15.49    \\
        $\sigma=0.50 \quad \mu=0.25 $  &   1.0s    &   11.65     \\  % Rs 3.297

  \bottomrule
    \end{tabular}
    \vspace{-5mm}
    \label{tab:noise}
\end{table}

% \subsection{Layer 2 Simulation}

% \begin{table}[h]
%     \centering
% %     \caption{Initial condition configuration in Simulation $1$}
% %     \begin{tabular}{lll}
% %         \toprule
% %         Parameters & Notations & Values                                                              80\\
% %         \midrule
% %         Initial Battery Temperature & $T_{bat,0}$ & $25^{\circ}C$ \\
% %         Initial Ambient Temperature & $T_{amb,0}$ & $25^{\circ}C$ \\
% %         \bottomrule
% %     \end{tabular}
% %     \label{tab:parameter}
% % \end{table}

% \paragraph{Simulation 2 - High initial temperature and ambient temperature} 

% \begin{table}[h]
%     \centering
%     \caption{Initial condition configuration in Simulation $2$}
%     \begin{tabular}{lll}
%         \toprule
%         Parameters & Notations & Values                                                                 \\
%         \midrule
%         Initial Battery Temperature & $T_{bat,0}$ & $-10^{\circ}C$ \\
%         Initial Ambient Temperature & $T_{amb,0}$ & $-10^{\circ}C$ \\
%         \bottomrule
%     \end{tabular}
%     \label{tab:parameter}
% \end{table}

% \subsection{UDDS Test}

% The microscopic traffic simulator VISSIM is used to simulate traffic on a flat
% single-lane roadway.

% \subsection{Hardware in the loop} To evaluate the potential capacity of the proposed speed optimization, we
% conducted a HIL test.

\section{Conclusion}

In this paper, we developed an optimization algorithm for dual-motor connected and automated electric vehicles, integrating vehicle dynamics control with powertrain-level optimization to achieve a comprehensive predictive control loop for the electric drivetrain to maximize energy savings. Simulation results demonstrate its effectiveness in optimizing speed control and motor torque allocation for traffic simulation scenarios developed from real-world data collected in Chattanooga, TN. The algorithm achieved a 12.80-24.52\% energy reduction, with torque allocation optimization of the dual motors contributing 3.16-4.40\% of the benefits. Additionally, the solution time for the implemented optimization algorithm remained within the MPC control update interval, providing a solid foundation for future laboratory testing and eventually real-world applications. For future work, we plan to incorporate a battery thermal management optimization system into the model, aiming to explore additional optimization potential over longer prediction horizons. Moreover, we will conduct laboratory testing to further validate the effectiveness of the implemented model.

% \begin{figure}[htbp]
%     \centering
%     \subfigure[front]{
%         \includegraphics[width=0.4\linewidth]{fig/OPveh_11084_Tf.epsc}
%         \label{fig:sub1}
%     }
   
%     \subfigure[rear]{
%         \includegraphics[width=0.4\linewidth]{fig/OPveh_11084_Tr.epsc}}
%         \label{fig:sub2}
    
%     \caption{Efficient map and operating for front and rear motors }
%     \label{fig:main}
% \end{figure}

% \subsubsection{Future Works}
% Even proposed approach considering the battery thermal management system,  model may not precise for some vehicle separate the battery and cabin thermal management system.   

% % \section{Conclusion}
% In this paper, we proposed a joint optimization on powertrain and thermal management of CAV. 

\bibliographystyle{ieeetr}
\bibliography{egbib}

\end{document}